\documentclass{emulateapj}
\def\gsim { \lower .75ex \hbox{$\sim$} \llap{\raise .27ex \hbox{$>$}} } 
\def\lsim { \lower .75ex \hbox{$\sim$} \llap{\raise .27ex \hbox{$<$}} } 
\newcommand{\vv}{$m_{606}$} 
\newcommand{\vi}{$m_{814}$} 
\shorttitle{The G1 Clump Overdensity in M31} 
\shortauthors{Faria et al.} 

\usepackage{amssymb}
\usepackage{amsmath}
\usepackage{amsbsy}
\usepackage{epsfig}
 
\begin{document} 
 
\title{Probing the Nature of the G1 Clump Stellar Overdensity in the Outskirts of M31\altaffilmark{1}} 
 
\author{Daniel Faria\altaffilmark{2},  Rachel A. Johnson\altaffilmark{3}, 
Annette M.~N. Ferguson\altaffilmark{4}, Mike J. Irwin\altaffilmark{5},  
Rodrigo A. Ibata\altaffilmark{6},  Kathryn V. Johnston\altaffilmark{7} 
Geraint F. Lewis\altaffilmark{8}, Nial R. Tanvir\altaffilmark{9}} 
\altaffiltext{1}{Based on observations made with the NASA/ESA Hubble  
Space Telescope, obtained at the Space Telescope Science Institute,  
which is operated by the Association of Universities for Research in  
Astronomy, Inc., under NASA contract NAS 5-26555. These observations are  
associated with program GO9458.} 
 
\altaffiltext{2}{Lund Observatory, Box 43, SE-221 00 Lund, Sweden; daniel@astro.lu.se} 
 
\altaffiltext{3}{Astrophysics, Oxford University, Denys Wilkinson Building, Keble Road, Oxford, OX1 3RH, UK} 
\altaffiltext{4}{Institute for Astronomy, University of Edinburgh, Royal Observatory Edinburgh, Blackford Hill, Edinburgh, EH9 3HJ, UK} 
\altaffiltext{5}{Institute of Astronomy, Madingley Road, Cambridge, CB3 0HA, UK} 
\altaffiltext{6}{Observatoire de Strasbourg, 11, rue de  
l'Universit\'{e}, F-67000, Strasbourg, France} 
\altaffiltext{7}{Astronomy Department, Wesleyan University, Middletown, CT 06459, USA} 
\altaffiltext{8}{Institute of Astronomy, School of Physics, A29,  
University of Sydney, NSW 2006, Australia} 
\altaffiltext{9}{Department of Physical Science, University of  
Hertfordshire, College Lane, Hatfield, AL10 9AB, UK}

\begin{abstract} 
 
We present deep HST/ACS observations of the G1 clump, a distinct 
stellar overdensity lying at $\sim 30$~kpc along the south-western 
major axis of M31 close to the G1 globular cluster (Ferguson et 
al. 2002). Our well-populated colour-magnitude diagram reaches $\sim$7 
magnitudes below the red giant branch tip with 90\% completeness, and 
allows the detection of various morphological features which can be 
used to derive detailed constraints on the age and metallicity of the 
constituent stellar population.  We find that the 
colour-magnitude diagram is best described by a population with a 
large age range ($\gtrsim 10$~Gyr) and a relatively high mean metallicity 
[M/H]$= -0.4$.  The spread in metallicity is constrained to be $\lesssim 
0.5$ dex.  The star formation rate in 
this region has declined over time, with the bulk of the stellar mass 
having formed $>6$~Gyr ago. Nonetheless, a non-negligible mass 
fraction ($\approx 10\%$) of the population has formed in the last 
2~Gyr.  We discuss the nature of the G1 Clump in light of these new 
stellar population constraints and argue that the combination of 
stellar content and physical size make it unlikely that the structure 
is the remnant of an accreted dwarf galaxy.  Instead, the strong 
similarity between the stellar content of the G1 Clump and that of the 
M31 outer disk suggests the substructure is a fragment of the outer 
disk, perhaps torn off from the main body during a past 
accretion/merger event; this interpretation is consistent with extant 
kinematical data.  If this interpretation is correct, our analysis of 
the stellar content provides further evidence that the outskirts of 
large disk galaxies have been in place for a significant time. 
 
\end{abstract} 
 

\section{Introduction} 
Within the currently-favoured cosmological framework, large spiral 
galaxies are thought to be assembled from the mergers and accretion of 
smaller building blocks and from the smooth accretion of gas. Under 
the assumption that at least some of the accreted satellites contain 
significant stellar components, signatures of this hierarchical 
assembly process are expected in the form of tidal debris and other 
stellar inhomogeneities.  Bullock \& Johnston (2005) have shown that 
such substructure is expected to be most readily visible in the far 
outer regions of galaxies. 
 
The Milky Way (MW) halo shows at least one unambiguous case of ongoing 
accretion (i.e. the Sagittarius dwarf, Ibata et al.\ 1994). Observing the MW 
halo is, however, not trivial; our location within the disk makes the 
identification and interpretation of halo structures through a 
dominant foreground disk population a challenge (e.g. Newberg et 
al. 2002; Juric et al.\ 2005).  Furthermore, depending on how 
membership is determined, various selection biases can enter into the 
construction of halo samples.  Studying halo populations in other 
galaxies is therefore an attractive alternative for testing ideas 
about galaxy assembly. 
 
The Andromeda galaxy (M31) has proven to be an excellent target for 
such studies. At a distance modulus of ($m-M$)$=24.47 \pm 0.07$ 
magnitudes (McConnachie et al.\ 2005), the M31 halo is well within 
reach for both photometric and spectroscopic investigations. Although 
the morphological type of M31 (Sb) is similar to that of the MW 
(SBbc), it has often been pointed out that many differences exist 
between the two systems. For example, M31 appears to have less gas 
than the MW (van den Bergh 2000) and shows a larger population of 
globular clusters, some of which are possibly younger than their MW 
counterparts (Fusi Pecci et al.\ 2005) (but see also Cohen et al.\ 
2005).  The recent discovery of a number of extended globular 
cluster-like objects in M31 (Huxor et al. 2005) further adds to the 
list of differences since similar such objects have not yet been found 
around the Milky Way. 
 
One of the first photometric studies of the resolved stellar 
populations in the outer regions of M31 was made by Mould \& Kristian 
(1986). Based on the colour of the brightest red giant branch stars, 
they found a more metal rich population than that of the MW 
halo. Since these first results, several subsequent studies have 
confirmed the presence of a population with a 
photometrically-determined metallicity peaking at [Fe/H] $\sim$ --0.6 
dex (e.g. Durrell et al.\ 1994, 2001; Holland et al.\ 1996; Bellazzini 
et al.\ 2003; Brown et al.\ 2003, 2006) and with a significant spread in metallicity. In 
comparison, the metallicity distribution of the MW halo population 
peaks at [Fe/H] $\sim$ --1.6 dex (Laird et al.\ 1988). 
 
Wide-field surveys of resolved populations provide the optimal means 
to probe the structure of nearby galaxies. The Isaac Newton Telescope 
Wide Field Camera survey has mapped M31 out to a radius of 55~kpc and 
beyond using resolved star counts that reach 2--3 magnitudes below the 
tip of the red giant branch (Ibata et al.\  2001; Ferguson et al.\ 2002; 
Irwin et al.\ 2005). A similar survey with Megacam on CFHT is now 
mapping the south-eastern quadrant of the galaxy from 50--150~kpc 
(e.g. Martin et al.\ 2006).  The resolved star count technique allows 
effective surface brightnesses as faint as $\Sigma_V \sim 32$ 
magnitudes per square arcsecond to be reached and has uncovered a 
wealth of substructure at large radius in M31.  In particular, 
Ferguson et al.\ (2002) found evidence for large-scale spatial and 
colour inhomogeneities in the galaxy outskirts, including a giant 
stellar stream (Ibata et al.\ 2001).  While the stream is almost 
certainly due to an ongoing accretion event, the nature of the other 
substructures around M31 is currently much less clear.  Recent 
studies of the stellar populations in the far outer regions of M31 have 
revealed what may be the M31 counterpart to the Milky Way halo. Both 
star counts and spectroscopy indicate that a metal-poor, 
pressure-supported power-law halo component begins to dominate at 
distances $>$ 30 kpc from the centre of M31 (Irwin et al.\ 2005, 
Chapman et al.\ 2006, Kalirai et al.\ 2006), whereas the photometric 
studies described above all lie within this radius.  The origin of this 
outer halo population, and its relationship to the inner halo 
population as well as the discrete substructures, has yet to be 
established. 
 
One of the most prominent features identified in Ferguson et 
al. (2002) is the G1 clump stellar overdensity shown in 
Figure~\ref{fig:pointing}.  Named for its proximity to the G1 globular 
cluster, the feature is centered at RA(2000)$=00^{h} 35^{m} 28^{s}$, 
Dec(2000)$=39^{\circ} 36^{\prime} 19^{\prime\prime}$ which corresponds 
to a projected radius of 29.6~kpc from the nucleus, almost directly 
along the south-western major axis. The feature spans more than $\sim$ 
10 kpc and has an estimated absolute magnitude of M$_{V}\approx -12.6$ 
(Ferguson et al.\ 2002). The close proximity to the G1 globular cluster 
($\approx 30'$ projected separation) has fueled 
speculation that the two might be physically related and motivated 
several recent studies.  Rich et al.\ (1996, 2004) present HST/WFPC2 
observations of the field population near G1 (though quite far from 
the center of the G1 clump, see Figure~\ref{fig:pointing}) and find 
that the cluster and neighbouring field exhibit rather different 
colour-magnitude diagram (hereafter, CMD) morphologies.  This argues 
against the overdensity being simply due to stars tidally-stripped 
from the cluster.  More evidence against such an association comes 
from examining the kinematics of the stars in the G1 clump overdensity 
and in the cluster (Reitzel et al.\ 2004; Ibata et al.\ 2005). Red giant 
stars in the G1 clump have radial velocities concentrated around 
$V_{helio} \approx -450~{\rm km~s}^{-1}$, which differs considerably 
from the velocity of the G1 globular cluster itself (--331 km s$^{-1}$ 
(Meylan et al.\ 2001)). 
 
To further investigate the nature and origin of the various 
substructures around M31, we have obtained deep photometry 
of ten fields using the Advanced Camera for Surveys (ACS) on board the 
HST during Cycles 11 and 13.  Preliminary results for six of these 
fields (including the G1 Clump) are presented in Ferguson et 
al. (2005). Our analysis revealed distinct morphological variations 
between many of the CMDs, indicating age and metallicity 
variations within the substructure. This finding strongly suggests 
that not all stellar overdensities in the outskirts of M31 are derived 
from a single accretion event. In this paper, we present a detailed 
analysis of the G1 clump field, deriving new constraints on the origin 
of this feature and exploring its possible relation to the M31 outer 
disk. 
 
\section{Observation and Data Reduction} 
 
The G1 clump field is one of eight fields in the outer regions of M31 
observed during Cycle 11 with the ACS Wide Field Camera as part of 
GO\#9458 (PI Ferguson).  Another two substructure fields were observed 
in Cycle 13 (GO\#10128). The ACS Wide Field Camera covers 
202$\times$202 arcsec on the sky which is equivalent to  
$\sim$0.8$\times$0.8 kpc at the distance of M31. 

\begin{figure*}[t!]
\noindent\includegraphics[width=39pc]{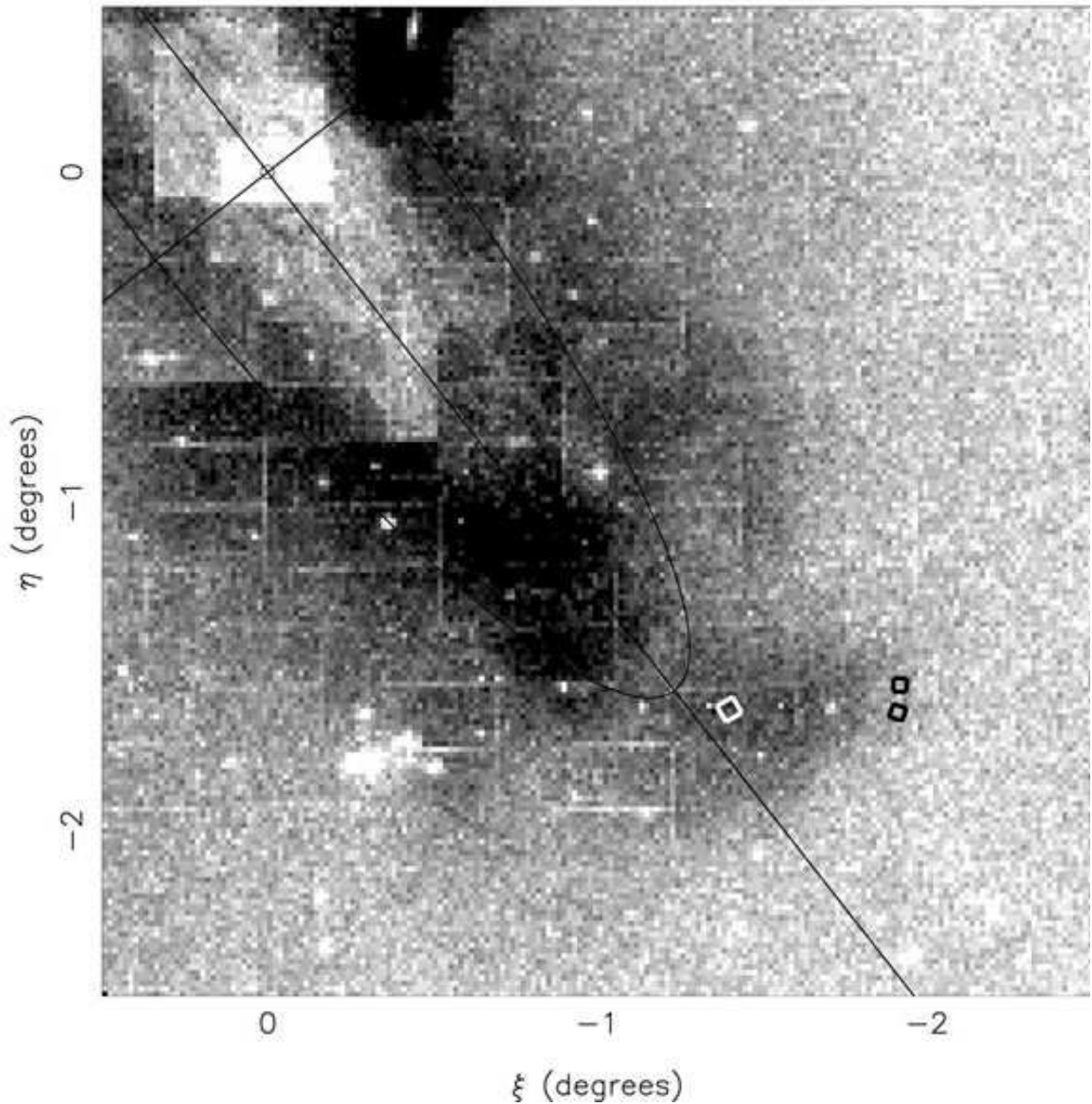}  
\figcaption{Surface density plot of individual stars along the 
   south-western major axis in M31. {\it Solid lines} show the minor 
   and major axis, and an ellipse representing an inclined disk with 
   $i$ = 77.5$^{\circ}$ and a radius of 2$^{\circ}$ ($\approx$ 27 
   kpc). The G1 clump is seen as an overdensity of stars just outside 
   the ellipse along the major axis at $\xi$ $\sim -1.5$ and $\eta$ 
   $\sim -1.75$. The {\it white box} indicates the position of our 
   HST/ACS field. The {\it upper black box} shows the position of the 
   HST/WFPC2 field (PID 5464), centered on the G1 globular cluster, 
   studied by Rich et al. (1996). The {\it lower black box} shows the 
   position of the HST/WFPC2 field (PID 9099) studied by Rich et 
   al. (2004).\label{fig:pointing}} 
\end{figure*}

The position of our ACS G1 clump field was selected to coincide with 
the peak of the stellar overdensity (see Figure~\ref{fig:pointing}). 
The field was observed for one orbit in the F606W filter (broad $V$) 
and two orbits in the F814W filter ($I$), all within a single 
visit. Integer pixel dithers were executed between multiple 
sub-exposures to aid warm pixel and cosmic ray rejection. Final 
exposure times were 2430 s in F606W and 5150 s in F814W.  We note that 
previous HST studies of this region have been based on fields lying at 
the very periphery of the stellar overdensity (Rich et al.\ 1996, 
2004). 
 
The images were first processed through the ACS pipeline and then 
combined per passband using Multidrizzle (Koekemoer et al.\  2002) 
within PyRAF. We used the default mode of Multidrizzle which 
calculates the offsets between dithered exposures using the 
information in the image header world coordinate system. We checked 
the accuracy of this method by measuring the alignment of the 
individual drizzled frames before the final combine operation.  We 
found images taken in a given filter to be registered to better than 
0.1 pixel. We also checked that the centers of stars were not being 
misidentified as cosmic rays, and no problems were noted. 
For the final drizzle, we used the Lanczos3 kernel with 
\texttt{pixfrac} and \texttt{scale} set to unity. 
 
Photometry was obtained using the IRAF implementation of DAOPHOT 
(Stetson 1987). The stellar density in the G1 field is low enough to 
allow precise results from aperture photometry alone. An aperture 
radius of 2 pixels was found to give the best results (i.e. the 
tightest CMD).  PSF-fitting photometry was also carried out in order 
to reject non-stellar objects such as background galaxies, and cosmic 
rays which were not previously eliminated.  A spatially-invariant PSF 
was created for each filter using the $\sim$150 brightest stars on the 
combined images.  After PSF-fitting, sources were retained in the 
final photometry list if their magnitude errors, sharpness and $\chi$ 
values all lay within 3-sigma of the average value at their magnitude. 
Final aperture photometry was then re-run on the cleaned star 
list. 
 
Aperture corrections were derived as follow. The corrections from 2 to 
4 pixel aperture radii were found from the average of several tens of 
bright stars on each frame. No significant variation in the 
aperture correction over the frame was found. In order to get good 
signal-to-noise, the aperture correction from 4 to 8 pixels was found 
using the PSF template star.  Finally the photometry was corrected 
from 8 pixel to `infinite' aperture using the tabulated values given 
in Table 3 in Sirianni et al.\ (2005). Table~\ref{tab:photcal} 
summarises the aperture correction values used. As a further 
test that there was no systematic variation of the photometry across the ACS field, 
we produced colour magnitude diagrams in different sub-regions of the chip. We 
found that these were identical, with no detectable variation in position or colour 
of the morphological features observed.

The photometry was placed on the Vegamag system using the values 
given in Table 10 of Sirianni et al.\ (2005).  We adopted a reddening in the G1 clump 
field of E($B-V$) = 0.06 (Schlegel et al.\ 1998).  The final Vegamag 
magnitudes used in this paper (referred to as \vv\ and \vi\ 
throughout) are then\\ 
 
\begin{displaymath} 
m_{\rm fil} = \rm -2.5log(m(2)) - C(2-\infty) + zpt - A_{\rm fil}, 
\end{displaymath} 
 
\noindent where m(2) is the measured value, in ADU/s, in the 2 pixel 
radius aperture, and the aperture correction, $C$, zeropoints, $zpt$, 
and extinction corrections, A$_{\rm fil}$, are given in 
Table~\ref{tab:photcal}. 
 
At a few points in the analysis, we require Johnson-Cousins magnitudes 
(referred to as $V$ and $I$) in order to be able to compare to models 
in the literature. These are obtained using the calibration presented in 
Sirianni et al.\ (2005). The transformation equations have the following form 
for ($V-I$) $>$ 0.4: 
 
\begin{displaymath} 
\rm V = m_{\rm 606OB} + 26.331 + 0.340(V-I) - 0.038(V-I)^{2} 
\end{displaymath} 
\begin{displaymath} 
\rm I = m_{\rm 814OB} + 25.496 - 0.014(V-I) + 0.015(V-I)^{2}, 
\end{displaymath} 
 
\noindent where m$_{\rm 606OB}$ = \vv \space -- 26.398 and m$_{\rm 
814OB}$ = \vi \space -- 25.501 (i.e. zero point subtracted 
magnitudes). $V$ and $I$ are obtained by iterating the equation above 
until convergence is reached ($\sim$ 10 iterations). An initial 
estimate of ($V-I$) = \vv \space -- \vi \space was used. As  
discussed in Sirianni et al.\ (2005), errors of a few percent can 
be introduced by making this transformation. 
 
Completeness tests were carried out separately for each filter.  The 
artificial stars used in the completeness tests were created using the 
PSFs described above. The completeness value for a given magnitude 
was calculated by adding approximately 1500 artificial stars with that 
magnitude to the image, and then running the same photometric pipeline 
as used for the data (e.g.\ detection, aperture photometry, PSF fitting 
and pruning, and final aperture photometry). The final completeness 
value is then the percentage of the input stars which are recovered. 
Completeness tests were carried out at a number of magnitudes spanning 
the entire magnitude range in the data.  Figure~\ref{fig:completeness} 
shows the completeness levels for the \vi \space and \vv \space 
magnitudes respectively. 

\begin{table}[t!] 
\begin{tabular}{lll}  \\
\hline 
\hline   
             & F606W & F814W \\ 
\hline 
aperture correction 2 - 4 & --0.22 & --0.31 \\ 
aperture correction 4 - 8 & --0.078 & --0.087 \\ 
aperture correction 8 - $\infty$  & --0.105 & --0.106 \\ 
zeropoint (Vegamag)       & 26.398 & 25.501 \\ 
extinction correction     & 0.173  & 0.117  \\ 
\hline 
\end{tabular} 
\caption{Aperture corrections and zeropoints used 
in photometric calibration}
\label{tab:photcal} 
\end{table}

\section{The Colour-Magnitude Diagram of the G1 Clump}   
 
Figure~\ref{fig:CMD} shows the  
CMDs of the G1 clump ACS field, which contain a total of 44420 stars.  The error 
bars show average $\pm$1 sigma errors on the magnitudes determined 
from the artificial star tests.  For magnitudes brighter than 
\vi$_{,0}$ $\sim$ 23.5 (\vv$_{,0}$ $\sim$ 24), the error bars are 
negligible.  The 90\% and 50\% completeness limits found from the 
tests are indicated by solid lines. 
 
The main features in the CMD can be summarised as follows: 

\begin{figure}[t!]  
\noindent\includegraphics[width=19pc]{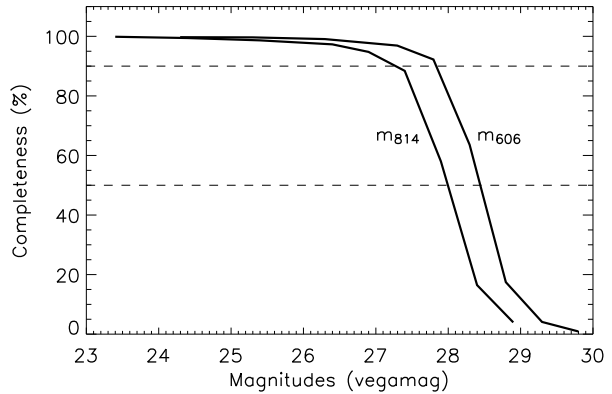} 
   \figcaption{Completeness in \vi \space and \vv \space derived from  
artificial star tests. { \it Dashed lines} indicate the 90\% and 50\%  
completeness levels.\label{fig:completeness}}     
\end{figure}

\begin{itemize} 
\item A broad, well-populated Red Giant Branch (RGB) of hydrogen-shell 
burning stars. The width is much larger than can be explained by the 
photometric errors alone, suggesting a spread in metallicity and/or 
age. CMDs of the outer regions of M31 have been known for a long time 
to show metal-rich RGBs with significant dispersion (e.g. Mould \& Kristian 1986; 
Durrell et al.\ 1994, 2001; Holland et al.\ 1996), and the RGB in the 
G1 Clump field is no exception. 
 
\item A pronounced Red Clump (RC) of core He burning stars at magnitude \vi$_{,0}$ 
$\sim$ 24.1 indicating a dominant population of stars with an intermediate 
age (i.e. 2 -- 10 Gyr). There is no obvious extended Horizontal Branch (HB), 
which suggests the absence of a significant population of truly ancient ($>10$~Gyr) 
stars. However, note that the faint blue HB expected for an ancient  
metal-poor population could be masked by the blue plume. 
 
\item A Blue Plume (BP) at (\vv \space -- \vi)$_{0}$ $\sim$ 0.1 extending from \vi$_{,0}$ 
  $\sim$ 26 to \vi$_{,0}$ $\sim$ 23, which is a possible signature of young 
  main-sequence stars. 
 
\item A clump of stars at \vi$_{,0}$ $\sim$ 23.0 and (\vv \space -- 
\vi)$_{0}$ $\sim$ 0.90 which we identify as the Asymptotic Giant 
Branch (AGB) bump, a feature caused by a temporary slowing down of 
luminosity increase at the beginning of a star's AGB phase. 
 
\item A low-level overdensity of stars just below the RC at \vi $\sim$ 
   24.7 and (\vv \space -- \vi)$_{0}$ $\sim$ 0.75. This is identified 
  with the RGB bump, a feature which is caused by the temporary drop 
  in an RGB star's luminosity when the expanding hydrogen-burning 
  shell reaches the chemical discontinuity left behind by the 
  convective layer during its main sequence phase. 
 
\end{itemize} 

\begin{figure*}[t] 
\epsscale{1.} 
\includegraphics[angle=90,width=17cm]{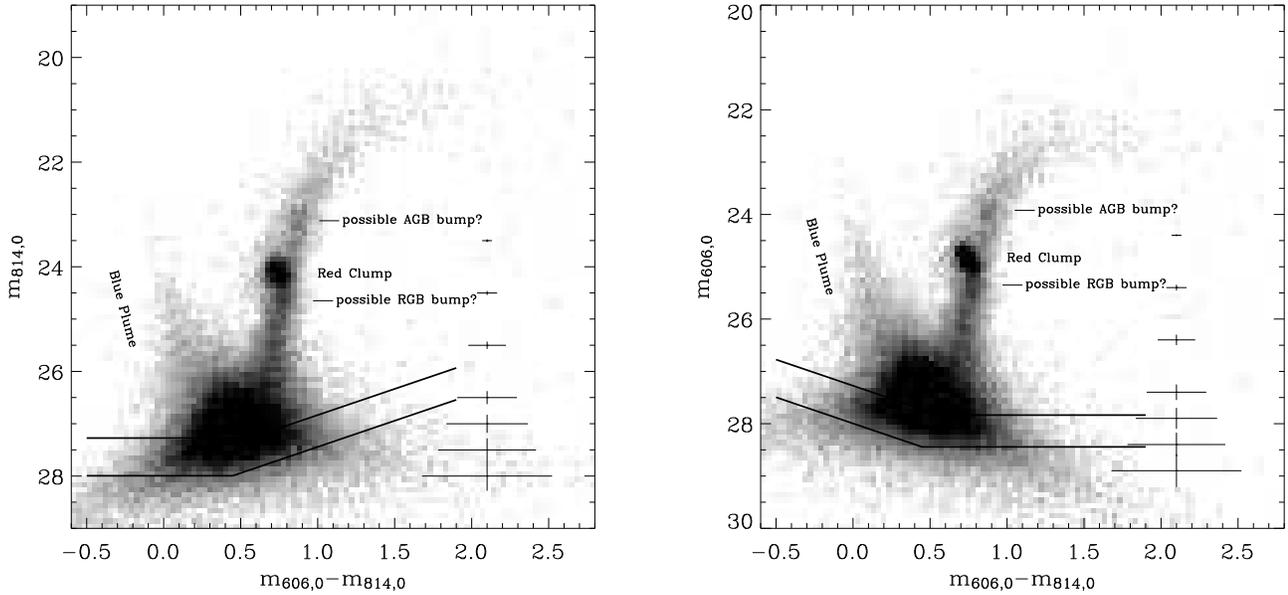} 
\figcaption{(\vv \space -- \vi)$_{0}$, \vi$_{,0}$ \space($left$) and (\vv  \space-- \vi)$_{0}$, \vv$_{,0}$ \space ($right$) 
Hess diagrams (44,420 stars) for our G1 clump field. The 
{\it solid lines} show the 90\% and 50\% completeness limit determined from 
artificial star tests. Typical errors along the red giant branch are plotted on the right hand 
side.\label{fig:CMD}} 
\end{figure*}

\section{Constraining the Age and Metallicity of the G1 Clump} 
 
Although our CMD is the deepest yet obtained for this region of M31, 
it does not reach the level of old main sequence turnoffs. As a 
result, we are unable to conduct a rigorous study of the full star 
formation and chemical evolution history of the G1 Clump. 
Nonetheless, our CMD is of sufficient quality and depth to allow a 
much more detailed exploration of the stellar populations than has 
been carried out in previous work.  Our adopted approach is to use 
initial isochrone fitting to guide our choice of model stellar 
populations and then explore how well different combinations of age 
and metallicity within this range can reproduce the observed CMD 
morphology and luminosity function in detail. 
 
Our modeling exploits the Girardi isochrones (Girardi et al.\ 2000) 
transformed into the ACS filter bandpasses. Metallicities are 
calculated assuming [M/H] = log(Z/Z$_\odot$), [$\alpha$/Fe] = 0 and 
Z$_\odot$ = 0.019.  The model CMDs used are created using the {\sc 
testpop} program from the StarFISH software developed by Harris \& 
Zaristky (2001). In short, {\sc testpop} creates artificial CMDs based 
on an input star formation history and data describing observational 
parameters such as completeness and photometric errors. A detailed 
description of {\sc testpop} is given in Harris \& Zaritsky (2001). To 
create the synthetic CMDs discussed below, we have assumed a distance 
modulus of $(m-M)$ = 24.47 (McConnachie et al.\ 2005), a binary 
fraction of 50\% and a Salpeter IMF slope of --1.35. The StarFISH 
software provides two ways to define completeness and errors, either 
by using an output table from the artifical star tests, or by fitting 
functions to the results of these tests.  We have used the second of 
these two methods.  For the error model, we fit an exponential 
function to the errors found in the artificial star tests. The 
completeness is modeled using two separate $arctan$ functions which 
characterise the completeness function above and below the 50\% 
completeness level. 
 
\subsection{The Blue Population in the G1 Clump} 
 
\subsubsection{The Blue Population as Young Main Sequence Stars} 
 
The most striking difference between our CMD and those previously 
presented for this region (Rich et al.\ 1996, 2004) is the clear 
presence of a BP at (\vv \space -- \vi)$_{0}$ $\sim$ 0.1 extending 
from \vi$_{,0}$ $\sim$ 26 to \vi$_{,0}$ $\sim$ 23. In the following, 
we investigate the BP under the assumption that it is a young 
population of main-sequence stars. 
 
\begin{figure}[t!] 
\noindent\includegraphics[width=19pc]{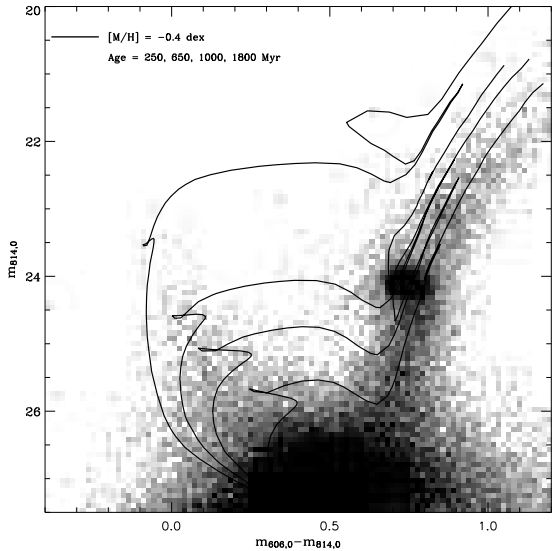} 
   \figcaption{(\vv \space -- \vi)$_{0}$, \vi$_{,0}$ \space Hess diagram with overlaid 
   Girardi et al. (2000) isochrones with [M/H] = --0.4 dex and ages = 250, 650, 1000, and 
   1800 Myr.\label{fig:BPiso}} 
\end{figure}

Figure~\ref{fig:BPiso} shows the BP and overlaid Girardi isochrones 
with [M/H] = --0.4 dex and ages = 250, 650, 1000, and 1800 Myr. The 
youngest isochrone with age = 250 Myr provides the best fit to the 
brightest stars, but lies blueward of the main plume, which seems 
better-described by somewhat older ages. More metal poor isochrones 
fall further towards the blue and hence fail to reproduce any aspect 
of the BP.  More metal rich isochrones fall further to the red, and 
could thus possibly be used to fit the redder part of the BP, however 
no single age and metallicity isochrone can fit the entire BP. This is 
the first suggestion of an age range -- implying extended star 
formation -- in this region of the CMD. 
 
\begin{figure}[t!]
\noindent\includegraphics[width=19pc]{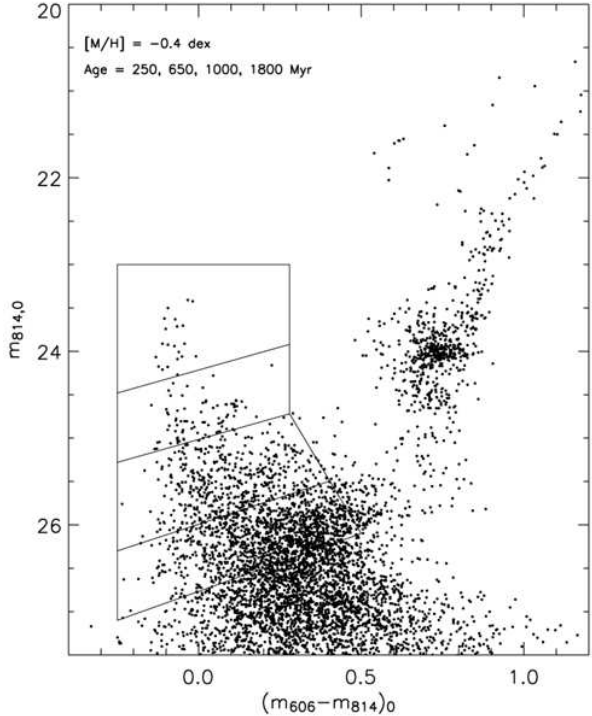} 
   \figcaption{(\vv \space -- \vi)$_{0}$, \vi$_{,0}$ \space synthetic CMD showing the model of the blue plume ([M/H] =  
--0.4 dex and ages = 250, 650, 1000, and 1800 Myr).\label{fig:BPmodel}} 
\end{figure}

To illustrate this, and to investigate the BP in more detail, we use 
synthetic CMDs produced with StarFISH. In Figure~\ref{fig:BPmodel}, a 
model CMD of a population with ages = 250, 650, 1000, and 1800 Myr and 
[M/H] = --0.4 dex is shown.  The four boxes indicated along the BP 
are used to roughly scale the number of stars in the four age 
components to the data in the following way: the brightest box only 
samples stars in the 250 Myr component, the second brightest box 
sample stars in the 250 and 650 Myr components, etc.  Using the 
brightest box to scale the 250 Myr component provides an estimate of 
the number of 250 Myr stars in the second box, which can then be used 
to scale the 650 Myr component, and so on. The total number of stars 
in the model BP is $\sim$5000, with increasingly more mass in older 
components (almost 65\% of the total mass is in the oldest component).  

The resulting model BP is a reasonable match to the data although 
still slightly too blue. No distinct subgiant branch can be seen for 
the younger components in the model, but one does appear for the 
oldest (and most dominant) population at \vi$_{,0}$ $\sim$ 26 and (\vv 
\space -- \vi)$_{0}$ $\sim$ 0.4. In the actual G1 Clump data, the BP 
merges with the RGB at a slightly fainter magnitude than this, and this may 
explain why a distinct turnoff is not discernible in the data. 
Our model of the BP is overly simplistic in that we have 
used only four ages to model the population. While this has been 
sufficient to demonstrate that an age range is required to match the 
observed population, it represents a star formation history consisting of 
discrete bursts.  In reality, the star formation history of the G1 
clump region is likely to have varied continuously.  Indeed, more 
sophisticated models which incorporate smaller age bins and extend to 
ages $>1800$ Myrs produce a smoother BP and the sub-giant branch seen 
in Figure~\ref{fig:BPmodel} disappears. However the scaling of such 
models becomes much more complicated, especially in the region where 
the BP merges with the RGB, and they are not developed further here. 

The model in Figure~\ref{fig:BPmodel} also predicts the CMD location 
of the evolved counterparts to such a young population.  Evolved stars 
from the two younger age components are expected to fall on the blue 
side of the RGB (i.e. at (\vv \space -- \vi)$_{0}$ \space $\leq$ 0.75 
and \vi$_{,0}$ \space $\leq$ 23 ), extending upwards from the observed 
RC. This feature is called a blue loop (see Gallart 1998 and 
references within). Although we do not observe a well-defined blue 
loop population in the G1 Clump CMD, we do see a number of stars in 
this region and their number is in good agreement with our simple 
model predictions. 
 
For the two older populations, ages = 1000 and 1800 Myr, the evolved 
stars are superimposed on the RGB and RC. The possible influence this 
will have on the interpretation of these features will be 
further discussed in the next section. We note, however, that the 
number of stars in our BP model is only $\approx 10\%$ of the total 
number of observed stars in the G1 clump field and that the bulk of these 
stars lie in the BP itself.

\subsubsection{The Blue Population as Blue Straggler Stars}\label{sec:bs} 
  
An alternative interpretation of the BP is that it represents a 
population of blue straggler stars.  CMDs of globular clusters and 
nearby dwarf galaxies often show a BP similar to the one we observe in 
the G1 clump. In globular clusters, where there is no recent star 
formation, the BP is identified with rejuvenated blue straggler stars, 
created either by stellar collisions or by mass transfer in primordial 
binaries (Davies et al.\ 2004). In more complex stellar systems, such 
as dwarf spheroidal galaxies, the nature of the BP is less clear. It 
has been attributed both to blue stragglers (e.g. in Ursa Minor, 
Carrera et al.\ 2002) as well as young and intermediate age populations 
(e.g. in Draco, Aparicio et al.\ 2001). 
 
In sparsely-populated stellar environments, such as the Milky Way
halo, the important mechanism for blue straggler formation is believed
to be mass transfer in primordial binaries (Carney et al.\ 2001). In
this scenario, one expects the blue stragglers to have masses lower
than twice the main-sequence turn-off mass ($M_{MSTO}$) of the
underlying population from which they have formed. However, in dense stellar
environments (i.e. globular clusters) theorists predict that
binary-binary collisions may occasionally produce blue stragglers
which are more massive than twice the turn-off mass (Fregeau et al.\
2004), but examples of this have yet to be observed. 
 
The two youngest isochrones overlaid in Figure~\ref{fig:BPiso} (ages = 
250 and 650 Myr) have main-sequence turn-off masses, $M_{MSTO}$ $\sim$ 
3.2 and 2.1 M$_\odot$.  A significant fraction of the stars on the BP 
thus appear to have masses $\geq$ 2 M$_\odot$ and reaching as high as 
3 M$_\odot$.  To explain such stars as blue stragglers formed via mass 
transfer, we would require a progenitor population of stars with 
$M_{MSTO} \gtrsim 1.5$ M$_\odot$.  Such a population 
would have ages of $\sim$ 2--3 Gyr . In Section 4.2, we will show that 
the G1 clump very likely contains such a population. We therefore 
cannot rule out blue straggler stars as contributors to the BP. 
 
We can further test the likelihood of a blue straggler contribution to 
the BP by calculating the blue straggler specific frequency, defined 
as the number of blue straggler stars compared to the number of RC 
stars, $S_{(BS)}$ = $N_{BS}$/$ N_{RC}$.  We calculate $N_{RC}$ by 
counting all stars on the RGB within 24.35 $<$ \vi$_{,0}$ $<$ 23.85, 
then correcting for the estimated number of RGB stars within this same 
region by measuring the number of stars above and below the RC. 
$N_{BS}$ was calculated by selecting all stars with \vi$_{,0}$ $<$ 26 
and (\vv \space -- \vi)$_{0}$ $<$ 0.5.  Using these definitions for 
$N_{BS}$ and $N_{RC}$, we find $S_{(BS)}$ = 1.7 for the G1 clump. This 
value can be compared to the values found for globular clusters 
($S_{(BS)}$ $\leq$ 2) and for the MW halo field population, $S_{(BS)}$ 
$\sim$ 4 (Preston \& Sneden 2000, and references within).  While the 
specific frequency of blue straggler stars in the G1 clump is slightly 
lower than that seen elsewhere, the discrepancy is not so significant 
that we can rule out this hypothesis. 
 
The blue straggler interpretation of the G1 Clump's BP population thus 
remains viable though we do not favour it here.  In particular, we 
note that the CMDs of other stellar overdensities in the outskirts of 
M31 do not show such strong BP populations as the G1 Clump (Ferguson 
et al.\ 2005), even though many of these other fields have similar or 
greater stellar densities.  This argues against the idea that the BP 
population is primarily composed of blue straggler stars, since such 
stars should be expected in similar amounts in all fields. 
Furthermore, as we will argue later, the existence of a genuinely 
young population is consistent with the location of the G1 Clump field 
within M31's HI disk and our derived metallicity for the population is 
consistent with that expected from extrapolation of M31's chemical 
abundance gradient. 
 
\subsection{The Old and Intermediate-Age Populations of the G1 Clump} 

\begin{figure}[t!] 
\noindent\includegraphics[width=19pc]{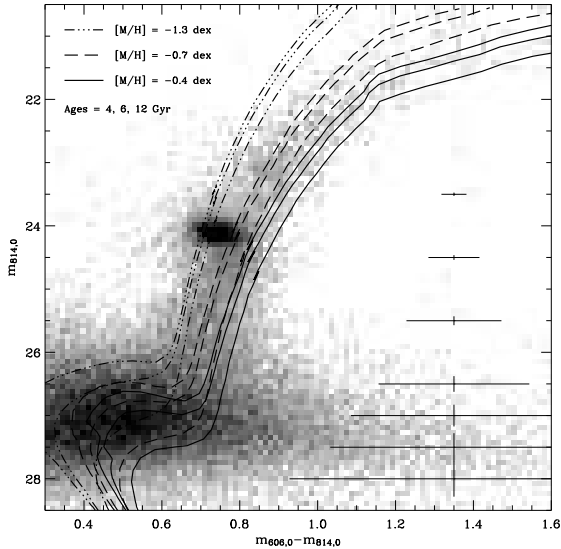} 
\figcaption{(\vv \space -- \vi)$_{0}$, \vi$_{,0}$ \space Hess diagram  
 of the red populations with overlaid Girardi et al. (2000) isochrones with [M/H]= 
--1.3, --0.7, and --0.4 dex and ages = 4, 6 and 12 Gyr.\label{fig:CMDiso}} 
\end{figure}

We can obtain constraints on the age and metallicity of old and 
intermediate-age populations of stars in the G1 Clump from analysis of 
the locus and width of the RGB in conjunction with the position and 
morphology of the RC (e.g.\ Ferguson \& Johnson 2001). In addition, the 
position of the AGB and the RGB bumps can be used to provide 
consistency checks on our best-fitting models (e.g. Alves \& 
Sarajedini 1999) provided that these low-level features can be 
correctly identified in the CMD. 
 
\subsubsection{Comparison to Theoretical Isochrones} 
 
As a first step, we compare the G1 Clump's RGB to the Girardi 
isochrones calculated for a range of ages and metallicities. 
Figure~\ref{fig:CMDiso} shows the CMD overlaid with isochrones of ages 
= 4, 6, and 12 Gyr and metallicities [M/H] = --1.3, --0.7, and 
--0.4 dex for each age. 
 
The metal poor isochrones ([M/H] = --1.3) all fall on the blue side of 
the RGB, and the curvature at the bright end is clearly not 
correct. The more metal rich isochrones provide better matches, 
although no single isochrone appears to fit the RGB along its whole 
length. At the faint end, the RGB locus is best matched by an 
intermediate age and metallicity model (i.e. 4--6 Gyr and 
[M/H] $\sim$ --0.7 dex), whereas at the bright end, more metal rich 
isochrones with [M/H] $\sim$ --0.4 dex do equally well. 
 
\begin{figure}[t!] 
\noindent\includegraphics[width=19pc]{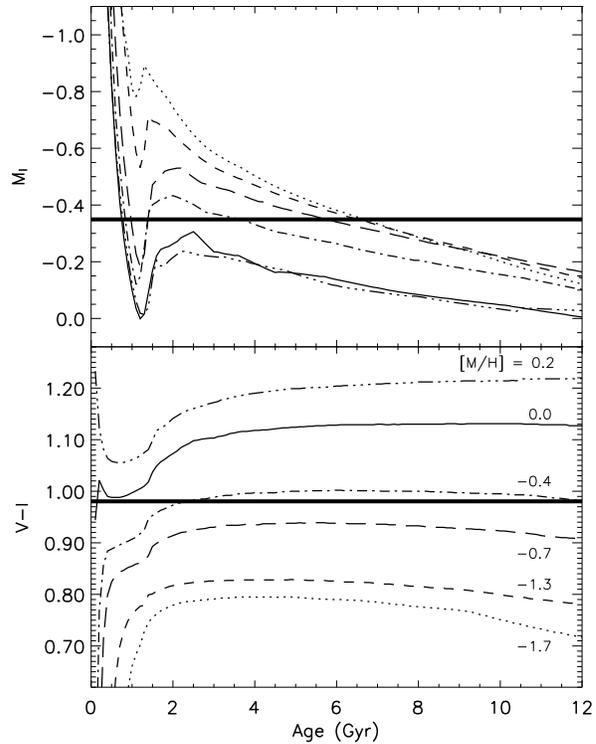} 
   \figcaption{Girardi \& Salari (2001) models of RC behavior as a function  
of age and metallicity. {\it Thick solid lines} indicate the position of the red  
clump in the G1 Clump field.\label{fig:RCmodels}} 
\end{figure} 

 \begin{figure}[t!]   
\noindent\includegraphics[width=19pc]{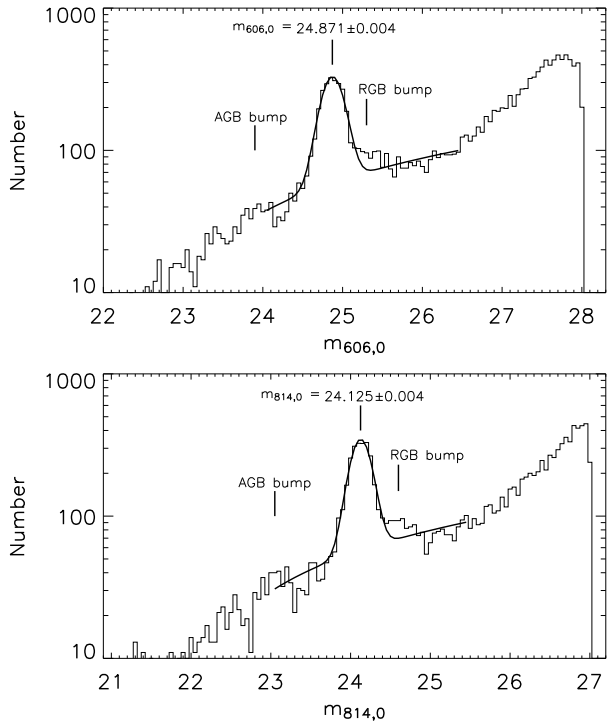} \figcaption{Luminosity function in 
\vv$_{,0}$ \space ({\it top}) and \vi$_{,0}$ \space ({\it bottom}). {\it Solid lines} show Gaussian fits to the red clumps. \label{fig:RCfit}} 
\end{figure}

The RGB is much wider than can be explained by the photometric errors, 
especially above the RC, suggesting that it cannot be explained by a 
single age and metallicity population. An estimate of the maximum 
possible metallicity spread can be found if we assume a single age 
population. From Figure~\ref{fig:CMDiso}, the metallicity spread 
required to explain the full width of the RGB is $\gtrsim$ 0.6 dex 
spanning from $\lesssim$ --1 dex to $\gtrsim$ --0.4 dex.  On the other 
hand, if we assume a single metallicity, we find that the required age 
spread needed to explain the full width of the RGB must be greater 
than the the entire age span of the overlaid isochrones in 
Figure~\ref{fig:CMDiso} (i.e. $\gtrsim 8$Gyr). 
 
As is well known, it is impossible to distinguish between a spread in 
metallicity and a spread in age by looking solely at the RGB. 
Fortunately, our CMD shows several additional features which can be 
used to break the age-metallicity degeneracy, the most prominent being 
the RC. 
 
\subsubsection{Red Clump Models} 
 
Figure~\ref{fig:RCmodels} shows models from Girardi \& Salaris (2001) 
which illustrate the RC behavior as a function of age and metallicity. 
For ages older than $\sim$ 2 Gyr, the mean colour of the RC has a 
strong dependence on metallicity but almost no dependence on age.  On 
the other hand, for a given metallicity, the mean magnitude of the RC 
depends strongly on age, so that an older population is expected to 
have a fainter RC than a younger one.  Together the mean magnitude and 
colour of the RC can provide insight into the age and metallicity mix 
of the constituent stellar populations (e.g. Ferguson \& Johnson 2001; 
Rejkuba et al.\ 2005).

The position of the RC in the G1 Clump CMD was found by fitting a 
Gaussian to the luminosity function in \vv$_{,0}$ \space and 
\vi$_{,0}$, as shown in Figure~\ref{fig:RCfit}. Following Paczynski \& 
Stanek (1998), a 5-parameter Gaussian was used with the two additional 
parameters representing a constant $y$-offset and linear slope term.   
The centroids of the Gaussians are: \vv$_{,0}$ \space = 
24.871 $\pm$ 0.004 and \vi$_{,0}$ \space = 24.125 $\pm$ 0.004, which 
gives (\vv \space -- \vi)$_{0}$ \space = 0.746 $\pm$ 0.006.  Using the 
transformations from the ACS photometric system to $V$ and $I$ 
described in Section 2, these values translate into $M_{I,\rm{RC}} = 
-0.349$, $M_{V,\rm{RC}} = 0.631$, and $(V-I)_{RC} = 0.980$, using ($m-M$)$=24.47$. 
Comparing these values with the models in Figure~\ref{fig:RCmodels} 
(solid lines), we infer a stellar population with a mean metallicity 
[M/H] $\sim$ --0.4 dex and intermediate age $\sim$ 3.5 Gyr.  These 
values are broadly consistent with those found earlier from isochrone 
comparisons.  We note there is some degree of degeneracy in the models in 
Figure~\ref{fig:RCmodels}; our RC properties could also be consistent 
with a $\sim$1-2 Gyr solar metallicity population.  This will be 
further commented on later in this section. Furthermore, uncertainties in 
the distance modulus could change the mean age derived by this 
relatively crude method by several Gyr. 
 
\begin{figure}[t!] 
\noindent\includegraphics[width=19pc]{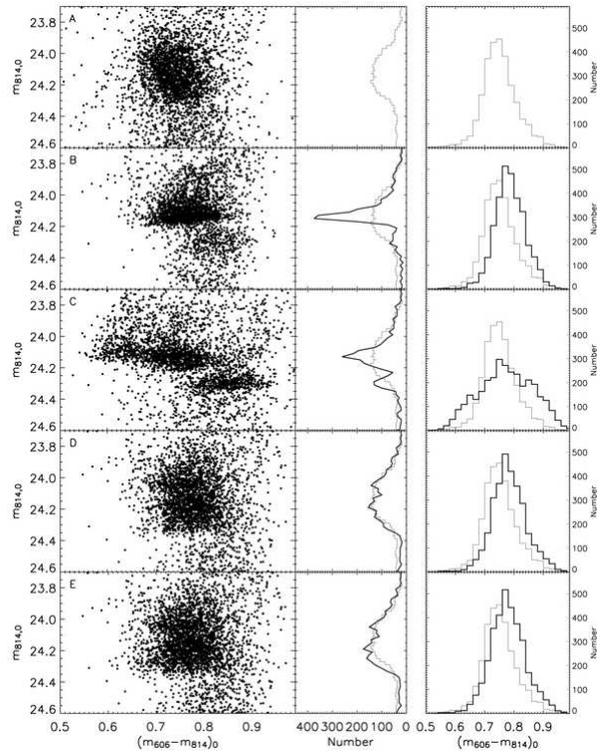} 
   \figcaption{Synthetic CMDs ({\it 
left}), luminosity functions ({\it center}), and colour distributions 
({\it right}) showing observed (light) and model (bold) 
distributions. (A) Shows the observed red clump of the G1 clump 
field. (B) shows a single age and metallicity model (age = 4 Gyr and 
[M/H] = --0.4 dex). (C) Shows a model with a significant metallicity 
spread (age = 4 Gyr, [M/H] = --1.3, --0.7, --0.4, and 0.0 dex). (D) 
Shows a model with a significant age spread and a single metallicity 
(age = 2, 4, 6, 8, and 10 Gyr, [M/H] = --0.4 dex). (E) Shows a model 
with a significant age spread and a small spread in metallicity (age = 
2, 4, 6, 8, and 10 Gyr, [M/H] = --0.4 dex, and age = 6, 8, and 10 Gyr 
for M/H] = --0.7 dex).\label{fig:RCmultiplot}} 
\end{figure}

The analysis above provides only a mean age and metallicity 
of the population, whilst in reality there could 
be several different populations contributing to the CMD.  We now 
proceed to use the detailed morphology of the RC to place contraints 
on the mix of populations that might be present. 
Figure~\ref{fig:RCmultiplot} shows a comparison between the observed 
RC and the RC of various synthetic populations spanning a range of 
ages and metallicities. For both the real data and the synthetic 
models, we show the CMD, luminosity function, and colour 
distribution for the RC region. The results from the data are 
shown in Figure~\ref{fig:RCmultiplot}A and overlaid on each of the models 
to aid with the comparison. The number of stars 
in the RC region of the model populations is adjusted to match the 
real RC to within a few percent. 

The first model (Figure~\ref{fig:RCmultiplot}B) is of a single age and 
metallicity population similar to the mean properties derived above 
(age = 4 Gyr and [M/H] = --0.4 dex). The model RC morphology is 
clearly different to the data -- the vertical extent of the RC model 
is much too small and slightly too red; the corresponding 
luminosity function emphasizes this even further. This comparison 
indicates that the observed RC cannot be fit by a single age and 
metallicity population. 

\begin{figure*}[t!] 
\epsscale{1.} 
%\plotone{f10.eps} 
\includegraphics[width=16cm]{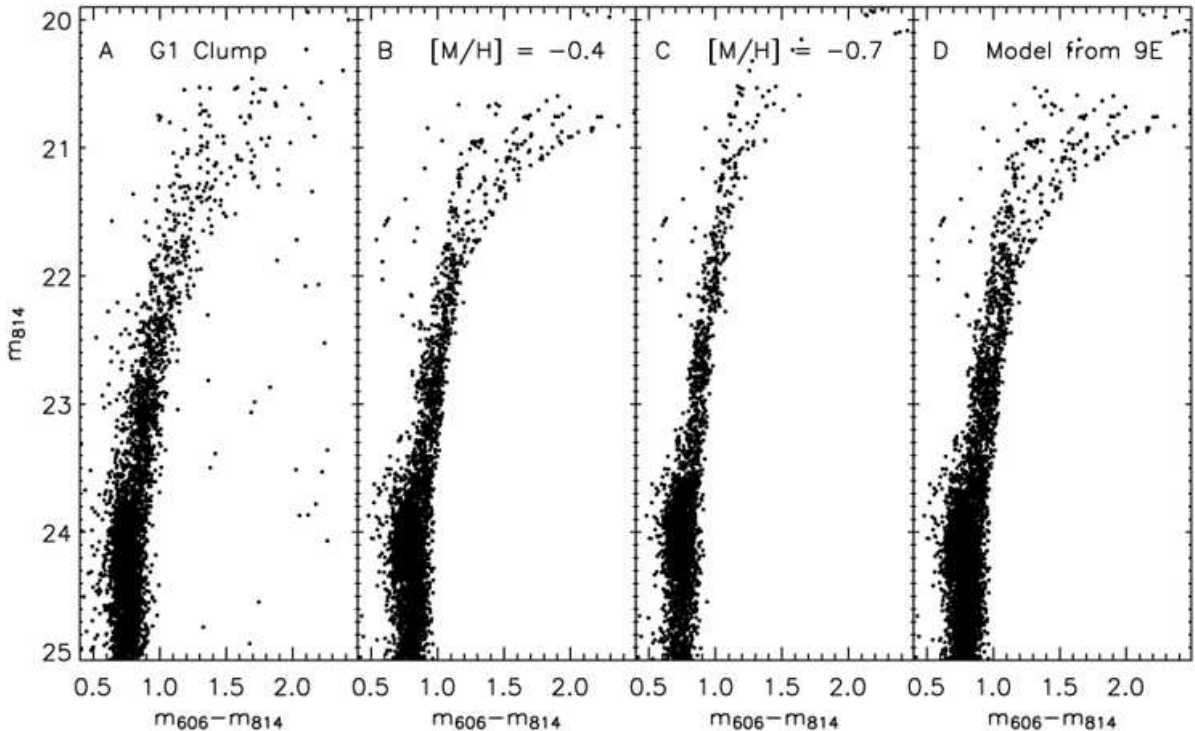} \figcaption{Synthetic (\vv 
   \space -- \vi)$_{0}$, \vi$_{,0}$ \space CMDs 
   showing the red giant branch. (A) Shows the red giant branch of the 
   G1 clump field. (B) Shows a model with a single metallicity and a 
   large age spread (age = 2, 4, 6, 8, and 10 Gyr, [M/H] = --0.4 
   dex). (C) Shows the same as (B) but with lower metallicity ([M/H] = 
   --0.7 dex). (D) shows the same model as in Figure~\ref {fig:RCmultiplot}E. The blue plume model from Figure~\ref{fig:BPmodel} is overlaid 
   on the models in (B), (C) and (D).\label{fig:RGB_models}} 
\end{figure*}

The model in Figure~\ref{fig:RCmultiplot}C has a single age but a 
spread in metallicity (age = 4 Gyr, [M/H] = --1.3, --0.7, --0.4, 
and 0.0 dex). This model produces a vertical extent which is similar 
to the data, but the morphology is clearly incorrect.  In addition to 
there being two distinct RC sequences in this model, the overall 
colour spread is much too large. In fact, it turns out that any 
significant population of stars with a metallicity higher than $\sim$ 
--0.2 dex or lower than $\sim$ --0.7 dex results in a RC with a colour 
spread that is larger than observed. On the other hand, restricting 
the metallicity spread to be $\sim$ --0.2 to $\sim$ --0.7 dex results 
in a RC with a vertical extent which is much too small compared to the 
observed spread in Figure~\ref{fig:RCmultiplot}A.  This comparison 
indicates that even a single age population with a range in 
metallicity fails to reproduce the observations.

Figure~\ref{fig:RCmultiplot}D shows a model with a pure age spread 
(age = 2, 4, 6, 8, and 10 Gyr) and a single metallicity ([M/H] = --0.4 
dex). The scaling of each age component was done to fit the luminosity 
function of the data in a similar way to the scaling of the BP model. 
The model RC provides a good match to the observations in terms of 
both general morphology as well as vertical extent. The colour 
of the model RC is slightly too red ($\sim$0.02 mag) indicating that 
the metallicity of the model is slightly too metal-rich, which is in 
good agreement with Figure~\ref{fig:RCmodels}. The metallicities of 
our models are limited by what is available in the Girardi isochrone 
set, so we are not able to model the effects of small metallicity 
variations.  Our modelling is not sophisticated enough to precisely 
constrain the age of the stellar population and there is some 
uncertainty in the oldest ages present. For example, almost identical 
RC morphologies are obtained for a slightly higher metallicity and a 
lower maximum age, or for slightly lower metallicities and a higher 
maximum age.  Nonetheless, our synthetic modeling of the RC 
demonstrates that an age spread is required to reproduce the 
observations.  This is further strengthened by a two-sided 
Kolmogorov-Smirnov test which yields a probability of $\sim$ 40\% that 
our model in Figure~\ref{fig:RCmultiplot}D and the data are identical. 
 
In order to test the effects of introducing a small metallicity 
spread, Figure~\ref{fig:RCmultiplot}E shows the same model as in 
Figure~\ref{fig:RCmultiplot}D but with the addition of a more 
metal-poor population in the older age bins (age = 6, 8 and 10 for 
[M/H] = --0.7 dex). The metal-poor component contains 25\% of the total mass in the model. We did not add stars in the younger age 
bins since they would produce an RGB which is significantly bluer than 
the observed RGB at the bright end (see Sect.~\ref{sect:RGB} and 
Figure~\ref{fig:RGB_models} ). As discussed above, such a metallicity 
range is still within that allowed by the colour width of the RC. The 
model RC is marginally wider in colour than the observed RC at the 
faint end and the agreement with the luminosity function is less good 
than before but this may be partially due to the discrete ages and 
metallicities used in our model.  It is possible that adopting a 
smooth age-metallicity relation across this metallicity range could 
provide an equally good match to the data as the model in 
Figure~\ref{fig:RCmultiplot}D.  In summary, although we cannot exclude 
a small metallicity range within the population, we have demonstrated 
that it is not required to model the observations of the red clump. 
 
As noted earlier, we expect the evolved counterpart of any young BP 
component to be superimposed on the RC (see Figures~\ref{fig:BPmodel} 
and \ref{fig:RCmodels}).  A more realistic model of the BP would have 
continuous star formation, instead of discrete ages, 
and since the change of the RC luminosity is a very strong function of 
age for ages less than $\sim$2 Gyr, the result would be that the stars 
from the younger population would be smeared out over $\sim$0.8 
magnitudes which is roughly the same vertical width as the old RC. We 
therefore believe that the presence of a very young ($<2$~Gyr) 
component in the RC would not significantly alter the conclusion 
derived from the above analysis. It is however possible that the small 
overdensity of stars below the RC that we identify with the RGB bump 
(see next section) could be somewhat contaminated by evolved stars in the young 
population or a mix of both. 
 
Also, we argued previously that if a population of age 2--3~Gyr was 
present, the BP could be interpreted as blue stragglers 
rather than a young population.  The modelling above shows that such a 
population is required to explain the morphology of the RC and thus 
that this scenario for the blue population cannot definitively be 
excluded. 
  
\subsubsection {Red Giant Branch Models} 
\label{sect:RGB} 
 
The modeling presented in the previous section demonstrates that a 
population with a large age spread and with either a single metallicity, or 
a small metallicity spread, can reproduce the overall properties of 
the RC. We next proceed to investigate whether the models explored in the 
red clump modelling can also reproduce the appearance 
of the RGB. Figure~\ref{fig:RGB_models} shows a 
comparison between the observed RGB and three different metallicity 
models, two with a single metallicity and a large age spread, and one 
with a large age spread and small metallicity spread. 
 
In Figure~\ref{fig:RGB_models}A, we show the observed G1 Clump 
RGB. Figure~\ref{fig:RGB_models}B and C show synthetic models of the 
RGB constructed with pure age spreads (age = 2, 4, 6, 8, and 10 Gyr 
for [M/H] = --0.4 and [M/H] = --0.7 dex, respectively). These 
metallicities were found to be the best matches in the RC modelling. 
In addition, our models include the contribution from the BP model 
shown in Figure~\ref{fig:BPmodel} (age$< 2$~Gyr).  As before, we find 
that the more metal-rich model ([M/H] = --0.4 dex) in 
Figure~\ref{fig:RGB_models}B shows a closer resemblance to the 
data. In particular, the youngest stars in the metal-poor 
model form a much straighter and bluer RGB at the bright end as 
compared to the observed CMD.  Furthermore, the [M/H] = --0.4 dex 
model shown here is a much better match to the lower part of the RGB 
than were the metal-rich isochrones overlaid in 
Figure~\ref{fig:CMDiso}. This is due to the younger age components now 
present in the model, which shift the RGB colour towards the blue, 
particularly at the fainter end. At brighter magnitudes, the evolved 
counterparts of the BP population scatter blueward of the main RGB. A 
BP model with a continuous star formation history would better 
populate the blue side of the entire RGB, similar to what is seen in 
the data. The single metallicity model in Figure~\ref{fig:RGB_models}B 
is slightly narrower than the data in the magnitude range 22.5 $<$ \vi 
$<$ 23.5. Figure~\ref{fig:RGB_models}D shows the RGB of the model with 
a large age spread and small metallicity spread introduced in 
Figure~\ref{fig:RCmultiplot}E. While the RC in this model did not 
look as good as in the single metallicity model, the RGB width in this 
model is more similar to the data, although the difference is small. 
 
We conclude that while the model with [M/H] = --0.4 dex and an 
age spread of 0.25--10~Gyr can adequately reproduce the overall width 
and shape of the observed RGB, a small metallicity spread, such as 
that shown in Figure~\ref{fig:RCmultiplot}E, could also be present.

\subsubsection{The RGB and AGB Bumps} 
 
In addition to the RC, two other overdensities of stars can 
be seen along the RGB at \vi$_{,0}$ \space $\sim$23.0 ((\vv \space -- 
\vi)$_{0}$ \space $\sim$0.9) and just below the RC at \vi$_{,0}$ 
\space $\sim$24.7 ((\vv \space -- \vi)$_{0}$ 
\space $\sim$0.75) and merging with the RC. Both features are 
indicated in Figure~\ref{fig:CMD} and Figure~\ref{fig:RCfit}. We 
identify these features with the AGB and RGB bump respectively. The 
brightnesses of these features are expected to vary with age and 
metallicity as shown in models presented by Alves \& Sarajedini 
(1999).  According to these models, the fact that the AGB and RGB bump 
straddle the RC in our data is a strong indication of metallicities 
above $\sim$ --0.7 dex. For a metallicity of [M/H] = --0.4 dex (the 
highest metallicity they examine), interpolating between the age bins 
given in their Table 2 leads one to expect $\Delta M_{V,\rm{ RGB \space 
bump - RC}}$ $\sim$0.4 mag for an 8 Gyr population, and $\sim -0.1$~mag 
for a 3 Gyr one.  Using the calibrations from Sirianni et al.\ and the 
distance modulus as before, we find the observed magnitude difference 
between the RGB bump and the RC to be $\Delta M_{V,\rm{ RGB \space 
bump - RC}}$ $\sim$0.4 mag.  Thus, the position of the RGB bump is in broad 
agreement with the the constraints derived earlier on the age and 
metallicity range of the population. 
 
Alves \& Sarajedini (1999) also predict the difference between the AGB 
bump and the RC, $\Delta M_{V,\rm{ AGB \space bump - RC}}$.  For [M/H] 
= --0.4 dex, they show that this value is expected to be $\sim$ -0.8 
magnitudes regardless of age. Since the RC luminosity varies with age, 
this implies that the peak of the AGB bump should be 0.8 magnitudes above 
the peak of the RC and should have a comparable vertical 
spread.  The presence of a low-level overdensity of stars comparable 
in luminosity extent to the RC at \vi$_{,0}$ \space $\sim$23, 
corresponding to $\Delta V_{\rm AGB \space bump - RC}$ $\sim$ -0.7, is 
thus consistent with the expected location of the AGB bump.

\subsection{The Final G1 Clump Model} 

\begin{table}[t!] 
\begin{tabular}{ccc} \\ 
\hline 
\hline 
 Age  & Mass & Fraction of Total Mass  \\ 
 (Gyr) &  (10$^{3}$ $\times$ M$_{\odot}$) & \% \\ 
\hline 
0.250 & 12.25 & 0.2 \\ 
0.650 & 25.00 & 0.5\\ 
1.0   & 45.00 & 0.8 \\ 
1.8   & 225.0 & 4.2 \\ 
2.0   & 300.0 & 5.5\\ 
4.0   & 800.0 & 14.8\\ 
6.0   & 1200 & 22.2 \\ 
8.0   & 1400 & 25.9 \\ 
10.0  & 1400 & 25.9\\ 
\hline 
\end{tabular} 
\caption{Stellar populations in our final model of the G1 clump.} 
\label{tab:model} 
\end{table}

Our final model of the G1 clump field is a combination of the BP model 
in Figure~\ref{fig:BPiso} and the RGB model with a pure age spread 
shown in Figure~\ref{fig:RCmultiplot}D and in 
Figure~\ref{fig:RGB_models}B.  This model contains populations of ages 
0.25--10~Gyr at a uniform metallicity of [M/H]$=-0.4$.  We stress 
again that our aim has not been to produce a high precision star 
formation history of the G1 Clump, but rather to come up with a simple 
model that can reproduce the detailed morphology of the CMD and thus 
yield insight into the nature of the constituent populations. We have 
previously commented on the fact that a small metallicity range could 
be present and still be consistent with the observed CMD.  Likewise, 
we have no way to constrain the presence of an additional component with 
ages $>10$~Gyr.  Figure 11 shows that our model CMD is in very good  
qualitative and quantitative agreement with the observations. 
Table~\ref{tab:model} shows the ages and masses in each component of 
the model; the bulk of the stars have ages $>6$~Gyr, however $\approx 
10$\% of the mass is contained in a population younger than 2~Gyr. 

\begin{figure}[t!] 
\noindent\includegraphics[width=19pc]{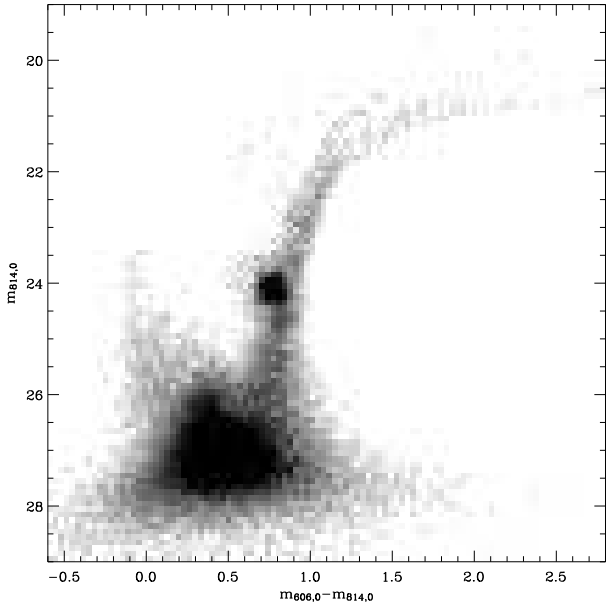} \figcaption{(\vv \space -- 
\vi)$_{0}$, \vi$_{,0}$ \space Hess diagram of the final 
model of G1 clump field (45,085 stars). The model is a combination of 
the blue plume model in Figure~\ref{fig:BPiso} and the red giant 
branch model with a pure age spread shown in  
Figure~\ref{fig:RCmultiplot}D and in 
Figure~\ref{fig:RGB_models}B. \label{fig:CMDfinalmodel}} 
\end{figure}

\begin{figure}[t!] 
\noindent\includegraphics[width=19pc]{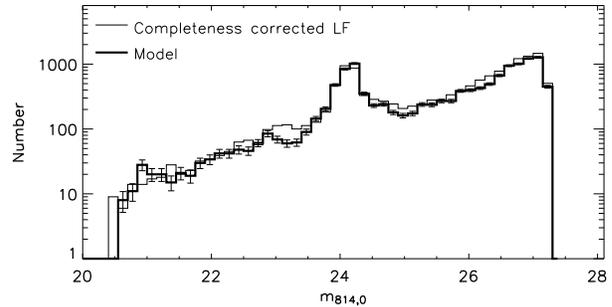} 
   \figcaption{Luminosity function in \vi$_{,0}$. The {\it thin line} shows the 
   completeness corrected G1 clump luminosity function for the 
   red giant branch. The {\it thick line} shows the corresponding luminosity function for our 
   final model shown in Figure~\ref{fig:CMDfinalmodel}. Poissonian 
   errors for each bin are shown on the model 
   luminosity function \label{fig:LFfinalmodel}} 
\end{figure}

Figure~\ref{fig:LFfinalmodel} shows the corresponding LF for the model 
compared to that observed for the G1 Clump. The model LF is in 
excellent agreement with the data along the entire RGB. A small 
deviation can be seen at \vi$_{,0}$ 
\space $\sim$ 23 where the model fails to reproduce in detail the AGB 
bump.  When more stars are added into the model, an AGB bump is 
produced at the correct position. The problem therefore is that the 
model AGB bump is weaker than the observed bump.  This problem was 
also seen in the modelling of Gallart (1998) 
 
For simplicity the analysis presented here has assumed no 
$\alpha$-enhancement. The effect of $\alpha$-enhancement has however been 
thoroughly investigated using the isochrones described in 
Salasnich et al.\ (2000). Introducing $\alpha$-enhancement has the effect 
of increasing the derived metallicity from [M/H] $\sim$--0.4 (non 
$\alpha$-enhanced) to [M/H] $\sim$0.0 ($\alpha$-enhanced), but does not 
alter the derived age spread. 
 
\section{Discussion and Conclusions} 
 
Our deep CMD analysis of the G1 Clump suggests a constituent 
population characterised by a relatively high mean metallicity 
([M/H]$\sim-0.4$) and an age range of at least 0.25--10 Gyr. In our 
model, the bulk of the stellar mass is of age $>6$~Gyr. Such a 
population represents a continuous, albeit declining, star formation 
history over at least 10~Gyr with only a mild amount of associated 
chemical evolution.  The presence of a young-to-intermediate age 
component in the vicinity of G1 was suggested in the WFPC2 studies of 
Rich et al.\ (2004).  Our deeper ACS CMD has provided conclusive 
evidence for this population and quantified its significance. 
  
The metallicity we derive for the bulk population in the G1 clump is 
significantly higher than values found in earlier studies.  Rich et 
al. (2004) find a metallicity of [Fe/H] $\sim$ --0.7 dex, which can be 
taken as [M/H] on the assumption of [$\alpha$/Fe]$=0$. We believe this 
difference can largely be explained by the presence of an age range 
within the G1 clump field.  If the metallicity is derived assuming a 
uniformly old stellar population (i.e. by comparison with Galactic 
globular cluster fiducials or old isochrones), a lower metallicity 
will be found. This is illustrated in Figure~\ref{fig:CMDiso} where it 
can be seen that a change in age from 12 Gyr to 4 Gyr will change the 
derived metallicity by $\sim$ 0.3 dex, or roughly the same difference 
between our value and that of Rich et al.\ (2004). 

The BP model in Figure~\ref{fig:BPmodel} can be used to estimate 
the total mass of the young population in the G1 Clump field. 
Assuming a Salpeter IMF with the exponent = \hbox{--1.35} and mass 
cut-off = 0.1 M$_\odot$, the total mass of our BP population in 
Figure~\ref{fig:BPmodel} is approximately 3 $\times$ 10$^5$ 
M$_\odot$. The 1800 Myr component comprises $\sim$65\% of the total 
mass. This translates to an average star formation rate of approximately 
10$^{-4}$ M$_\odot$/yr over the last 1.8 Gyr. 
 
\subsection{The G1 Clump as the Remnant of an Accreted Dwarf Galaxy} 
 
The morphology of the G1 clump -- i.e. an irregular lump distinct from 
the main disk (see Figure 1) -- suggests it could be the remnant of an 
accreted dwarf galaxy.  Recent studies of RGB kinematics in this 
region have shown the G1 Clump stars to have radial velocities similar 
but not identical to those of the rotating HI disk in M31 (Reitzel et 
al. 2004; Ibata et al.\ 2005).  Although this fact has been used to 
argue against the idea that the G1 Clump is an accreted dwarf, it is 
possible that some satellites could be 
accreted on prograde coplanar orbits.  Indeed, some argue that the 
stellar overdensity seen in the Milky Way towards Canis Major may be 
an example of such an accretion event (e.g. Martin et 
al. 2004). 
 
It is therefore worthwhile investigating whether the stellar 
populations of the G1 Clump could be consistent with a low mass 
satellite.  At first glance, the G1 clump appears rather different to 
present-day dwarf spheroidal galaxies around the Milky Way and M31, most 
of which show evidence for dominant old and metal-poor populations 
(see e.g. Dolphin et al.\ 2005; Da Costa et al.\ 2002).  The estimated 
absolute magnitude of the G1 clump is $M_{V}$ $\simeq$ --12.6 
(Ferguson et al.\ 2002).  Local group dwarfs of this magnitude 
typically have --2 $<$ [Fe/H] $<$ --1.5 (Mateo 1998), much more 
metal-poor than the G1 clump for which we find metallicity [M/H] 
$\sim$ --0.4 dex. 
 
There are, however, a few examples of dSph galaxies which contain both 
young-to-intermediate age stars as well as stars with moderate 
metallicities. One of these is the Fornax dwarf which contains a stellar 
population of mean age 5.4~Gyr (including stars as young as 0.2~Gyr) 
and with metallicities reaching to almost solar (e.g. Savianne et al.\ 2000; 
Tolstoy et al.\ 2006).  The other system is the Sagittarius dwarf 
galaxy, which the Milky Way is currently in the process of 
accreting. Analysis of the stellar content in the core of this galaxy 
reveals a dominant population of age $\sim 6$~Gyr (but lacking any 
very young stars) and metallicity [M/H]$=-0.5$ (e.g. Bellazzini et 
al. 1999; Monaco et al.\ 2005). In spite of the similarities in stellar 
content, the G1 Clump remains distinct from these systems in terms of 
its size and surface brightness.  At $\sim 10$~kpc, it is many times 
larger than the core diameters of either Fornax or Sagittarius and 
roughly 4 magnitudes fainter in peak surface brightness.  It thus does 
not easily fit into the sequence of properties which characterize 
known dwarf galaxies.  An additional argument against the dwarf galaxy 
hypothesis has been raised by Ibata et al.\ (2005). They point out that 
the extent ($\sim 10$~kpc) and velocity dispersion of 30 km/s of the 
G1 Clump would require a very considerable associated dark mass ($\sim 
10^9$~M$_{\odot}$) in order to be bound. Although we cannot completely 
exclude this hypothesis with existing data, we conclude that the G1 
Clump is unlikely to be an accreted dwarf galaxy. 
 
\subsection{The G1 Clump as Perturbed Outer Disk} 
 
Another hypothesis for the G1 Clump is that it is a perturbation of 
the M31 outer disk.  Our analysis of stellar populations in the G1 
Clump supports this idea on several grounds.  First of all, we have 
demonstrated that the G1 Clump has experienced continuous low-level 
star formation for most of the last 10~Gyr.  This implies a constant 
supply of gas, consistent with a location in the outer disk of M31. 
Indeed, Carignan et al.\ (2006) have shown that M31's HI disk extends 
out to at least 35~kpc, well beyond the location of the G1 Clump.  The 
recent average star formation rate of 10$^{-4}$ M$_\odot$/yr 
compares favourably with that determined for a location at 24~kpc in 
the outer disk by Williams (2002)(see his Figure 5(a)).  In addition, 
Williams (2002) finds that the outer disk is characterised by a 
declining star formation rate, consistent with what we 
have found for the G1 Clump. 
 
Further support for this picture comes from comparing the metallicity 
of the M31 gas disk with the metallicity of stars in the G1 Clump. The 
mean metallicity inferred for the G1 Clump, [M/H]$\sim-0.4$ dex, is 
rather high compared to that typically found in the outer regions of 
present-day gas disks (e.g. Ferguson et al.\ 1998). At face value, this 
might suggest that the G1 Clump cannot have formed from the gas disk 
in M31. M31 is unusual, however, in having a high mean gas-phase 
metallicity and a rather flat gradient (e.g. Trundle et al.\ 2002). At 
a radius of 26~kpc, the mean oxygen abundance in the disk as 
determined from HII regions is 40\% solar (Ferguson \& Urquhart, in 
preparation). In order to compare the gas-phase [O/H] and the stellar 
[Fe/H] abundances, we need to a adopt a value for [O/Fe].  Trundle et 
al. (2002) have compiled [O/Fe] and [$\alpha$/Fe] for supergiant stars 
in M31 and find that, within the errors, these values are consistent 
with being zero across the disk. This implies that 
[M/H]$=$[Fe/H]$\sim$[O/H]$\sim-0.4$ in the outer disk, and hence 
that stellar metallicity of the G1 Clump is in excellent agreement 
with the extrapolation of the M31 disk abundance gradient.  While it 
may not strictly be correct to compare the present-day metallicity of 
the gas disk with that of stars which formed several Gyr ago, the low 
star formation rate inferred for the G1 Clump implies that, in the 
absence of gas flows, little chemical evolution will have taken place 
in this region, even over long periods. This is supported by the small 
spread in metallicities inferred from our modelling.  We note that the 
high mean metallicity inferred for this region indicates a 
significant amount of pre-enrichment before the onset of star formation. 
 
Finally, the overall morphology of the G1 Clump shows some 
similarities (e.g. RGB and RC morphology) to several other fields that 
lie at large radius along the major axis in M31 (e.g. Ferguson \& 
Johnson 2001; Ferguson et al.\ 2005) suggesting that all may sample the 
same structure. Although the strength of the BP is observed to vary 
between some of these fields, this may simply reflect local variations 
in the recent star formation rate.  For example, the field observed by 
Ferguson \& Johnson (2001) shows a fainter and sparser BP than the G1 
Clump field despite lying at a similar distance along the major 
axis. On the other hand, the Ferguson \& Johnson (2001) field lies 
along the north-eastern major axis in a region where the gas disk 
severely warps away from the disk plane. As a result, the recent star 
formation history of this field may well be different from outer disk 
fields which lie within the main gas disk.  On the other hand, the 
fact that the old and intermediate-age populations in these two fields 
share similar properties may provide further evidence that the outer 
regions of the M31 disk have been in place for a significant time, 
contrary to some theoretical expectations (e.g. Abadi et al.\ 2003). 
 
If the G1 Clump does represent the outer disk, we require an 
explanation for its appearance as a distinct overdensity detached from 
the main body of M31.  N-body simulations indicate that the outer 
regions of disks can suffer significant damage during minor mergers 
(e.g. Quinn et al.\ 1993; Walker et al.\ 1996).   In these simulations, 
the final disk possesses distinct lumpiness at large radius, 
consisting of stars which existed in the initial stellar disk as well 
as stars stripped off the accreted satellite.  Another possible 
scenario is a large merger, similar to that studied by Springel \& 
Hernquist (2005). They show that the remnant of a major merger between 
similar-sized gas disks could still retain some disk structure.  In 
this scenario, the outer disk will consist of stars formed after the 
merger in the resettled gas disk.  The oldest stars in the outer disk 
would therefore date the merger event.  In the case of the G1 clump, 
we have shown the oldest stars are at least 8-10 Gyr and hence any 
significant merger of this sort must have happened a long time ago. 
The existence of copious substructure in the outskirts of M31 strongly 
supports the idea M31 has experienced at least one significant 
accretion event (Ferguson et al.\ 2002) and hence the hypothesis that 
the G1 Clump represents torn off disk appears viable.  Given the 
stellar population constraints derived here and the kinematical 
constraints derived elsewhere (Reitzel et al.\ 2004; Ibata et 
al. 2005), we strongly favour this hypothesis for the origin of the G1 
Clump.

\acknowledgements 
 
We thank Leo Girardi for providing isochrones in the ACS bandpasses 
ahead of publication. DF thanks Ivo Saviane for many valuable discussions. 
AMNF is supported by a Marie Curie Excellence Grant from the European 
Commission under contract MCEXT-CT-2005-025869. DF acknowledges the 
ESO Studentship Programme. Support for program GO9458 was provided by 
NASA through a grant from the Space Telescope Science Institute, which 
is operated by the Association of Universities for Research in 
Astronomy, Inc.  under NASA contract NAS 5-26555.

\newpage 
 
%TABLES 

%FIGURES 

\end{document}